\documentstyle[preprint,eqsecnum,aps,epsf,psfig]{revtex}
\newif\iftightenlines\tightenlinesfalse
\tightenlines\tightenlinestrue
\begin{document}

\title{Tau Neutrino Fluxes from Atmospheric Charm}
\author{L. Pasquali and M. H. Reno}
\address{
Department of Physics and Astronomy, University of Iowa, Iowa City,
Iowa 52242}

\maketitle

\begin{abstract}

We present an evaluation of the atmospheric tau neutrino flux  
in the energy range between $10^2$ and $10^6$ GeV. 
The main source of tau neutrinos is from charmed particle production
and decay.
The $\nu_\tau N\rightarrow \tau X$
event rate for a detector with a water equivalent
volume of 1 km$^3$ is on the order of 60-100 events per year
for $E_\tau>100$ GeV, reducing to 18 events above 1 TeV. 
Event rates
for atmospheric muon neutrino oscillations to tau neutrinos are also
evaluated.

\end{abstract}

\section{INTRODUCTION}

Recent measurements of the atmospheric neutrino flux by the Super-Kamiokande
Collaboration\cite{sk,superk} show a deficit of muon neutrinos in comparison
to theoretical predictions, while the 
measured electron neutrino flux is consistent with theory
assuming that all neutrino masses vanish. 
Earlier, lower statistics experiments already showed inconsistencies with
theoretical atmospheric flux predictions \cite{imb}.
On the basis
of event rates and the zenith angle dependence of the  muon neutrino deficit,
the Super-Kamiokande Collaboration has 
shown that their results could be explained by
neutrino oscillations between $\nu_\mu$ and $\nu_\tau$ \cite{superk}.
Oscillations imply at least one non-zero neutrino mass.
Definitive evidence of massive neutrinos 
requires modifying the standard model of electroweak interactions.

The Super-Kamiokande Collaboration measures neutrino fluxes from 
observations
of electrons and muons  in neutrino nucleon
interactions: $\nu_l + N\rightarrow l + X$.
In view of the importance of the question of whether or
not neutrinos have mass, one would like to see not
just $\nu_\mu$ disappearance, but also the $\nu_\tau$ appearance
coming from $\nu_\mu\rightarrow \nu_\tau$ oscillations.
Oscillation sources of $\nu_\tau$'s include
oscillations on the terrestrial scale
from atmospheric $\nu_\mu$'s as well as oscillations over
large astronomical distances of $\nu_\mu$'s produced in, for example,
active galactic nuclei \cite{agn}. 
A background to the flux of neutrinos from $\nu_\mu\rightarrow \nu_\tau$
are tau neutrinos produced directly
in the atmosphere.

Tau neutrinos are produced in the atmosphere by cosmic ray collisions 
with nuclei in the atmosphere, 
which produce charm quark pairs. A fraction
of the time, the emerging hadrons are $D_s$'s, which have a leptonic
decay channel
$D_s\rightarrow \tau \nu_\tau$
with a branching ratio of a few percent. 
The subsequent $\tau$ decays also contribute
to the atmospheric $\nu_\tau$ flux. Heavier mesons contribute to the
flux of tau neutrinos, but as we show below, they are negligible compared
to the $D_s$ contribution.

In 
this letter,
we outline the procedure to calculate the atmospheric tau neutrino flux.
The details of the method as applied to atmospheric electron neutrino,
muon neutrino and muon fluxes from charm decay appear in Refs. \cite{us}
and \cite{tig}.
We present our flux results
for the neutrino energy range of $10^2-10^6$ GeV, 
followed by the resulting $\nu_\tau + N
\rightarrow \tau +X$ event rates. For tau energies above 100 GeV, the
rate is on the order of $60-100$ events per year per km$^3$ water
equivalent volume. With a 1 TeV threshold, there are on the order of
20 events. We also evaluate the expected event rate for the tau neutrino
flux coming from $\nu_\mu\rightarrow \nu_\tau$ oscillations based
on a range of parameters consistent with the Super-Kamiokande results
\cite{superk}. For tau energies above a few hundred GeV, the atmospheric
tau neutrino background flux from $D_s\rightarrow \nu_\tau \tau$
dominates the tau neutrino flux from atmospheric muon neutrino oscillations.

\section{TAU NEUTRINO FLUX CALCULATION}

The main source of atmospheric tau neutrinos is the leptonic decay of the 
$D_s$: $D_s\rightarrow \tau\nu_\tau$, followed by 
$\tau\rightarrow \nu_\tau X$. For relativistic particles, a semianalytic,
unidimensional
approximate solution to cascade equations describing proton, meson and
lepton fluxes is a reliable approximation \cite{book,lipari,tig}.
The solutions rely on factorizing source terms in the cascade equations
into factors which are weakly dependent on energy times the incident
cosmic ray flux, here approximated by a proton flux. The source term
for $p\,$Air$\rightarrow D_s$, for a $D_s$ of energy $E$ 
and column depth $X$ as measured from the top of the atmosphere is
\begin{eqnarray}
S(p\rightarrow D_s)& \simeq & {\phi_p(E,X)\over \lambda_p(E)}
\int_E dE_p
{\phi_p(E_p,0)\over \phi_p(E,0)}{\lambda(E)\over \lambda_p(E_p)}
{dn_{p\rightarrow D_s}\over dE}(E;E_p)\\ \nonumber 
&\equiv & {\phi_p(E,X)\over \lambda_p(E)} Z_{pD_s}(E) \ .
\end{eqnarray}
Here $\phi_p(E,X)$ is the flux of cosmic ray protons
at column depth $X$. At the top of the atmosphere ($X=0$), following
Ref. \cite{tig}, we set
\begin{equation}
\phi_p(E,0)=1.7\ (E/{\rm GeV})^{-2.7}\ {\rm cm}^{-2}{\rm s}^{-1}
{\rm sr}^{-1}{\rm GeV}^{-1}, 
\end{equation}
valid for energies values lower than $5 \cdot 10^6$ GeV.
In Eq. (2.1), $\lambda_p$ is the proton interaction length and
$dn/dE$ is the cross section normalized energy distribution of the $D_s$
emerging from the proton-Air collision.
The quantity $Z_{pD_s}(E)$ is called a $Z$-moment. Generically, $Z$-moments
describe sources of particles of energy $E$, whether by production,
decay or energy loss through scattering. A complete discussion of the 
$Z$-moment method of solution appears in Refs. \cite{book} and 
\cite{lipari}. Its recent application to atmospheric muon, muon neutrino and
electron neutrino fluxes from charm decays is found Refs. \cite{tig}
and \cite{us}.

Solutions to the cascade equations in terms of $Z$-moments have two separate
forms, depending on whether the decay lengths are short compared to the
height of production (``low energy'') or long (``high energy'').
For $\nu_\tau$'s from $D_s$ (and $\tau$) decays, we confine our attention
to the neutrino energy range $10^2-10^6$ GeV. These neutrino
energies are  well below the critical
energy of $\sim 10^8$ GeV, above which 
decay lengths of the relativistic $D_s$'s and $\tau$'s
are longer than the vertical distance to height of production. 
Tau neutrinos are called ``prompt'' in the low energy regime. The
approximate solution for the $\nu_\tau+\bar{\nu}_\tau$ flux,
at the surface of the Earth, is
\begin{equation}
\phi_{\nu_\tau}(E) = {{Z_{pD_s}(E) Z_{D_s\nu_\tau}(E)}\over {1-Z_{pp}(E)}}
{\phi_p(E,0)} \ .
\end{equation}
The prompt tau neutrino flux is isotropic.
$Z_{pp}(E)$ accounts for the proton energy loss in proton-air collisions.
For $Z_{pp}(E)$, we use the results obtained by Thunman et al. in
their recent evaluation \cite{tig} using the Monte Carlo PYTHIA \cite{pythia}.
A similar, energy independent value was used in Ref. \cite{lipari}.

For $Z_{pD_s}(E)$, there are several approaches. Here we show the results
from next-to-leading order 
(NLO) perturbative QCD production of charmed quark pairs, scaled by a
factor of 0.13 to account for the fraction of $c\rightarrow D_s$
\cite{review}.
Details of the NLO calculation in the context of the prompt muon, muon
neutrino and electron neutrino fluxes appear in Ref. \cite{us}.
A second evaluation relies on Thunman et al.'s \cite{tig} $Z_{pD^0}(E)$,
rescaled by the ratio of $D_s$ to $D^0$ production taken to be 0.25.

The $D_s\rightarrow \nu_\tau$ decay $Z$-moments have several contributions.
The most straightforward is the direct $D_s\rightarrow \nu_\tau$ in the
two body decay, where \cite{book,lipari}
\begin{equation}
Z_{D_s\nu_\tau}^{(2\ body)}(E)=\int_0^{1-R_{D_s}} {dx\over x}
{Z_{pD_s}(E/x))\over Z_{pD_s}(E)}{\sigma_{pA}(E/x)\over \sigma_{pA}(E)}
{\phi_p(E/x,0)\over \phi_p(E,0)} {B\over 1-R_{D_s}}
\end{equation}
in terms of  
$x={E}/{E_{D_s}}$, $R_{D_s}=m_\tau^2/m_{D_s}^2$ and $B=0.043$, the 
branching ratio for $D_s\rightarrow \tau\nu_\tau$ \cite{gg}. We use
the inelastic proton-air cross section 
\begin{equation}
\sigma_{pA}(E)=\bigl(290-8.7\ln (E/{\rm GeV})+1.14\ln^2(E/{\rm GeV})
\bigr)\ {\rm mb} 
\end{equation}
parameterized in Ref. \cite{mielke}.

For neutrinos produced in the chain decay $D_s\rightarrow \tau 
\rightarrow \nu_\tau$ the decay $Z$-moment is given by
\begin{eqnarray}
Z_{D_s\nu_\tau}^{(chain)}(E) & = &
\int_0^1{dy\over y}\int_{R_{D_s}}^1 {dx\over x}
{Z_{pD_s}(E/(xy))\over Z_{pD_s}(E)}{\sigma_{pA}(E/(xy))\over \sigma_{pA}(E)}
{\phi_p(E/(xy),0)\over \phi_p(E,0)}\\ \nonumber
& & \quad \cdot
{B\over 1-R_{D_s}}
{dn_{\tau\rightarrow \nu_\tau}\over dy}
\end{eqnarray}
where $B$, $R_{D_s}$ and $\sigma_{pA}$ are the same as in the previous case 
while now $x={E_\tau}/{E_{D_s}}$ and $y={E}/{E_\tau}$. The $y$ distribution 
that appears in Eq. (2.6) can be generically written in the 
following way\cite{book}
\begin{equation}
{dn_{\tau\rightarrow \nu_\tau}\over dy} = B_\tau
\Bigl(g_0(y) - P_\tau(x) g_1(y)\Bigr)  
\end{equation}
where $B_\tau$ is the branching ratio for $\tau\rightarrow \nu_\tau X$ 
and $P _\tau (x)$ is the tau polarization. We have evaluated
the functions $g_0(y)$ and 
$g_1(y)$ for $\tau\rightarrow \nu_\tau \rho$, $\tau\rightarrow \nu_\tau \pi$,
$\tau\rightarrow \nu_\tau a_1$
\cite{bal} and $\tau\rightarrow \nu_\tau \ell \nu_\ell$.
They   
are collected in Table I, as are the branching fractions that 
we use. The purely leptonic decay also appears in Refs. 
\cite{book,lipari}. 
In terms of the energy of the parent $D_s$ ($E_{D_s}$), the tau 
polarization can be written as
\begin{equation}
P_\tau = {2 R_{D_s}\over 1-R_{D_s}}{E_{D_s}\over E_\tau } - {(1+R_{D_s})\over 
(1-R_{D_s})} \ .
\end{equation}
The total $D_s\rightarrow \nu_\tau$ $Z$-moment is
\begin{equation}
Z_{D_s\nu_\tau}=Z_{D_s\nu_\tau}^{(2 \ body)}+ Z_{D_s\nu_\tau}^{(chain)}\ ,
\end{equation}
where $Z^{(chain)}_{D_s\nu_\tau}$ includes the sum over all the tau decay
modes in Table I.

\section{TAU NEUTRINO FLUX RESULTS}

Using NLO perturbative QCD with $m_c=1.3$ GeV, and factorization scale
$M$ equal to twice the renormalization scale $\mu=m_c$ with the
CTEQ3 parton distribution functions \cite{ctq}, we obtain
$Z_{pD_s}$  between $10^{-5}$ and $10^{-4}$.
The Thunman et al. PYTHIA results (TIG) \cite{tig} give instead
$Z_{pD_s}$ between $3.5 \cdot 10^{-5}$ 
and $6.5 \cdot 10^{-5}$. The corresponding decay moments $Z_{D_s\nu_\tau}$
are 
$10^{-3}-5.5\cdot 10^{-3}$  and $0.8 \cdot 10^{-3}-
4.8 \cdot 10^{-3}$, respectively, over the energy range $10^2-10^6$ GeV.

The results for the tau neutrino fluxes scaled by $E^3$ are shown in Fig. 1 
where the solid line represents the flux obtained using NLO perturbative 
QCD, and the dashed line represents the flux obtained using the TIG $D_s$ 
production $Z$-moment.
In principle, one should also take into account the neutrinos produced by 
the semileptonic decay of the $b$ quark ($b\rightarrow c \tau\nu_\tau$) 
followed by the decay of the tau, but this contribution is
less than a few percent of charmed meson source.
The dotted 
line represents the flux from $b$ quark production and 
decay in the limit of zero 
polarization for the tau. 

The fact that the flux calculated using NLO 
perturbative QCD differs from the TIG PYTHIA based calculation
\cite{tig} is apparently due to the inclusion 
of fragmentation effects in the PYTHIA Monte Carlo.
A detailed comparison of the two approaches appears in Ref. \cite{us}.
The tau neutrino flux evaluated using NLO perturbative QCD can be 
parametrized as
\begin{equation}
\log_{10} \bigl(E^3 \phi_{\nu_\tau}(E)/({\rm GeV}^2/{\rm cm}^2 {\rm s\ sr}) 
\bigr)=-A+Bx-Cx^2-Dx^3
\end{equation}
where $x=\log_{10}(E/{\rm GeV})$, $A=7.08$, $B=0.765$, $C=0.00346$ and 
$D=0.00349$. The TIG rescaled result calculated here has
$A=6.69$, $B=1.05$, $C=0.150$ and $D=-0.00820$.

\section{DISCUSSION}

Large underground experiments may detect $\nu_\tau$-nucleon 
charged current 
interactions  $\nu_\tau +N\rightarrow \tau+X$. 
In the absence of $\nu_\mu
\rightarrow \nu_\tau$ oscillations, event rates \cite{gqrs} for $E_\tau>
100$ GeV and $E_\tau>1$ TeV are shown in Table II. We have summed over
neutrino and antineutrinos and
assumed that
there is no attenuation of the $\nu_\tau$ flux due to passage through
the Earth at these energies \cite{gqrs}. Detection of atmospheric
$\nu_\tau$'s in the absence of oscillations requires large detector volumes,
on the order of 1 km$^3$ water equivalent volume for $\sim 60-100$ events
per year with a tau energy threshold of 100 GeV.
Such large detector volumes for the detection of $\nu_\mu$ charged current
interactions are being considered \cite{antares}. 

In the case of neutrino oscillations, atmospheric muon neutrinos may oscillate
as they travel to the detector. Even with a small fraction of
$\nu_\mu\rightarrow \nu_\tau$ conversions, the copious production of
$\nu_\mu$'s in the atmosphere make this a potentially large source of
$\nu_\tau$'s. Again, assuming no attenuation due to 
neutrino interactions in the Earth, and approximating the height of 
production in the atmosphere by a constant 20 km
\cite{depth}, we have evaluated the
probability of oscillation in terms of neutrino energy and zenith angle. 
Using the Bartol muon neutrino flux \cite{bartol} from $\pi$ and $K$ decays
extrapolated to $E=10^6$ GeV, adding the NLO prompt $\nu_\mu$ flux \cite{us}
and integrating over zenith angle, Table
III shows the event rates for two limiting values of $\Delta m^2$, the 
difference between the squares of the tau neutrino and muon neutrino masses,
quoted by the Super-Kamiokande Collaboration \cite{superk}. We have set 
the oscillation mixing angle
$\sin^2(2\theta)=1$. For the larger mass difference,
$\nu_\mu\rightarrow \nu_\tau$ events overwhelm the prompt $\nu_\tau$ events
for $E>100$ GeV, while for the lower mass difference the rates are comparable.
Tau
neutrino oscillations to $\nu_\mu$
in this parameter range have a negligible depletion  of the
prompt $\nu_\tau\rightarrow \tau$ event rate.

For $E_\tau>1$ TeV, because prompt fluxes decrease less rapidly with energy
than the muon neutrino flux coupled with oscillations, the prompt 
$\nu_\tau$'s dominate the event rate.
The cross over to prompt $\nu_\tau$ domination of the flux occurs at
a few hundred GeV.
Another distinction between prompt
and oscillation tau neutrinos is the angular distribution. The prompt
source is isotropic, while the oscillation to $\nu_\tau$'s will be strongly
angular dependent, with the most $\nu_\tau$'s traveling upward because
of the long path length.

We have concentrated here on atmospheric sources of prompt $\nu_\tau$'s and
atmospheric $\nu_\mu\rightarrow \nu_\tau$. There are many flux models
of astrophysical sources of $\nu_\mu$'s \cite{agn}. If neutrinos have mass,
because of the large astronomical distance scales, on the order of half
of the $\nu_\mu$'s would oscillate to $\nu_\tau$'s. 
The detectability of
$\nu_\tau\rightarrow \tau$ conversions in the PeV energy
range has been explored
in the literature \cite{pakvasa}. 
It is clear from our calculation that
the atmospheric tau neutrino flux is negligible at PeV energies.

\acknowledgements
Work supported in part
by National Science Foundation Grant No.
PHY-9802403.

\begin{table}
\caption{Functions $g_0$ and $g_1$ in the tau neutrino energy distribution 
from $\tau$ decays, in terms of $y=E/E_\tau$ and $r_i=m_i^2/m_\tau^2$ and 
relative branching ratios.}
\begin{tabular}{lccc}
Process & $B_\tau$ & $g_0$ & $g_1$ \\ \hline
& & & \\
$\tau\rightarrow \nu_\tau \mu \nu_\mu$ & 0.18  &  ${5\over 3} - 
3y^2+{4\over 3}y^3$
& ${1\over 3} - 3y^2+{8\over 3}y^3 $\\
& & &\\
$\tau\rightarrow \nu_\tau \pi$ & 0.12 & ${1\over 1-r_\pi}$ 
& $-{2y -1+r_\pi\over (1-r_\pi)^2}$ \\
& & &\\
$\tau\rightarrow \nu_\tau \rho$ & 0.26 & ${1\over 1-r_\rho}$ 
& $-\biggl({2y-1+r_\rho\over (1-r_\rho)^2}
\biggr)\biggl({1-2r_\rho\over 1+2r_\rho}
\biggr)$\\
& & &\\
$\tau \rightarrow  \nu_\tau a_1$ & 0.13 & ${1\over 1-r_{a_1}}$ 
& $-\biggl({2y-1+r_{a_1}\over (1-r_{a_1})^2}
\biggr)
\biggl({1-2r_{a_1}\over 1+2r_{a_1}}
\biggr)$\\
& & & \\
\end{tabular}
\end{table}

\begin{table}
\caption{Charged current event rate per year
per km$^3$ water equivalent volume from the prompt $\nu_\tau+\bar{\nu}_\tau$
flux.
}
\begin{tabular}{lcc}
Threshold & NLO QCD & TIG 
 \\ \hline
100 GeV & 58 & 98 \\
1 TeV & 18 & 18 \\
\end{tabular}
\end{table}

\begin{table}
\caption{Charged current event rate per year
per km$^3$ water equivalent volume from $\nu_\mu+\bar{\nu}_\mu
\rightarrow \nu_\tau+\bar{\nu}_\tau$ oscillations,
assuming $\sin^2(2\theta)=1$.
}

\begin{tabular}{lcccc}
Threshold & $\Delta m^2=5\cdot 10^{-4}$ eV$^2$ & 
$\Delta m^2=6\cdot 10^{-3}$ eV$^2$ \\ \hline
100 GeV &  71 & 9100 \\
1 TeV &  0.036 & 5.2 \\
\end{tabular}
\end{table}

%figure 1
\begin{figure}
\psfig{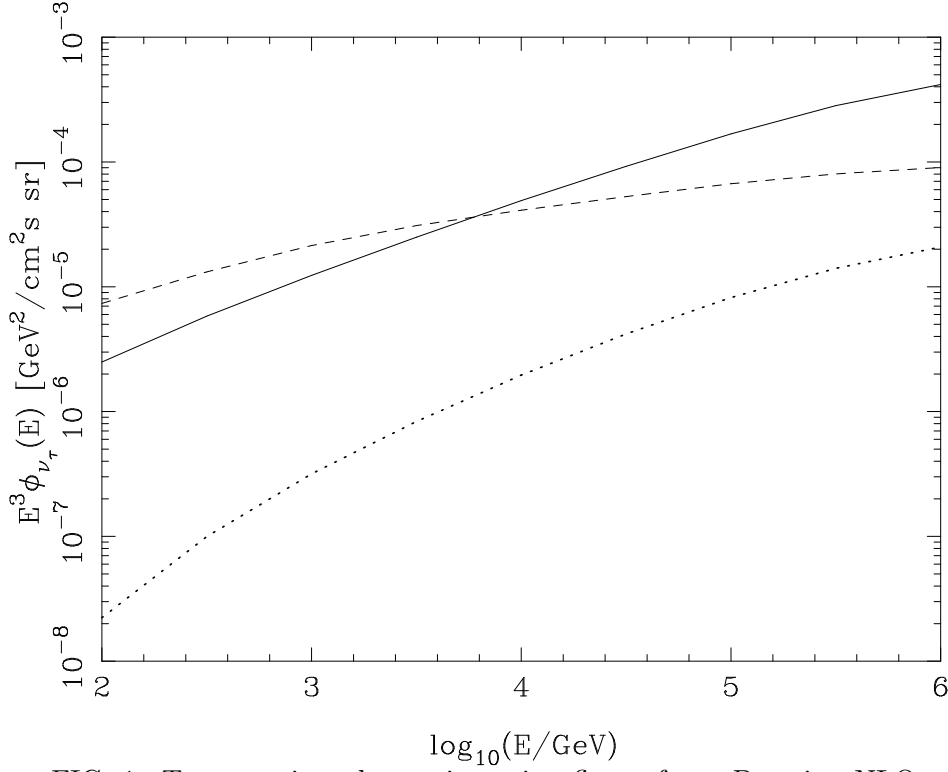}
\caption{Tau neutrino plus antineutrino
fluxes from $D_s$ using NLO perturbative QCD 
(solid line) and the TIG rescaled $Z$-moment (dashed line).
The tau neutrino plus antineutrino
fluxes from $b$ quark NLO perturbative QCD production and decay 
with the tau polarization equal to zero is shown by the 
dotted line.}
\end{figure}

\end{document}